 \documentclass[aps,prb,showpacs,groupedaddress]{revtex4}

\usepackage{graphicx}
\usepackage{bm}
\usepackage{subfig}
\begin{document}

\preprint{}
\title{ Plasmons for a two-Dimensional Array and a Bundle of nanotubes}
\author{Tibab McNeish }
\email{tmcneish@hunter.cuny.edu}
\author{Godfrey Gumbs}
\email{ggumbs@hunter.cuny.edu}
\author{Antonios  Balassis}
\email{abalassi@hunter.cuny.edu}
\affiliation{Department of Physics and Astronomy,
Hunter College at the City University of New York, \\
695 Park Avenue New York, NY 10065}
\date{\today}

\begin{abstract}
     We calculate the plasma excitations in a bundle as well as a two-dimensional (2D)
periodic array of aligned parallel multishell nanotubes on a substrate. The carbon nanotubes are oriented
perpendicular to the substrate. The model we use for the  system  is an electron gas confined to the surface of an
infinitely long cylinder embedded in a background dielectric medium. Electron tunneling between
individual tubules is neglected.   We include the Coulomb interaction between electrons on the same tubule
and on different tubules for the same nanotube and neighboring nanotubes. We present a self-consistent field
theory for the dispersion equation for intrasubband and intersubband plasmon excitations. For both the bundle
and 2D array of aligned parallel nanotubes, the dispersion relation of the collective modes is determined by
a three-dimensional wave vector with components in the direction of the nanotube axes and in the transverse
directions. The dispersion equation is solved numerically for a singlewall nanotube 2D array as well as a
bundle, and the plasmon excitation energies are obtained as a function of wave vector.  The intertube Coulomb
interaction couples plasmons with different angular momenta $m$ in individual nanotubes, lifting the $\pm m$
degeneracy of the single-nanotube modes. This effect is analyzed numerically as a function of the separation
between the tubules. We show that the translational symmetry of the lattice is maintained in the plasmon
spectrum for  the periodic array, and the plasmon energies have a periodic dependence on the transverse wave
vector $\bf q_{\bot}$. For the bundle, the Coulomb interaction between nanotubes gives rise to optical plasmon excitations.
\end{abstract}

\pacs{73.20.Mf,\ 73.61.Wp, \ 71.45.Gm, 61.46.+w}

\maketitle

\section{Introduction}
\label{sec1}

\begin{figure}[p]
\includegraphics[width=4.5in,height=5.6in]{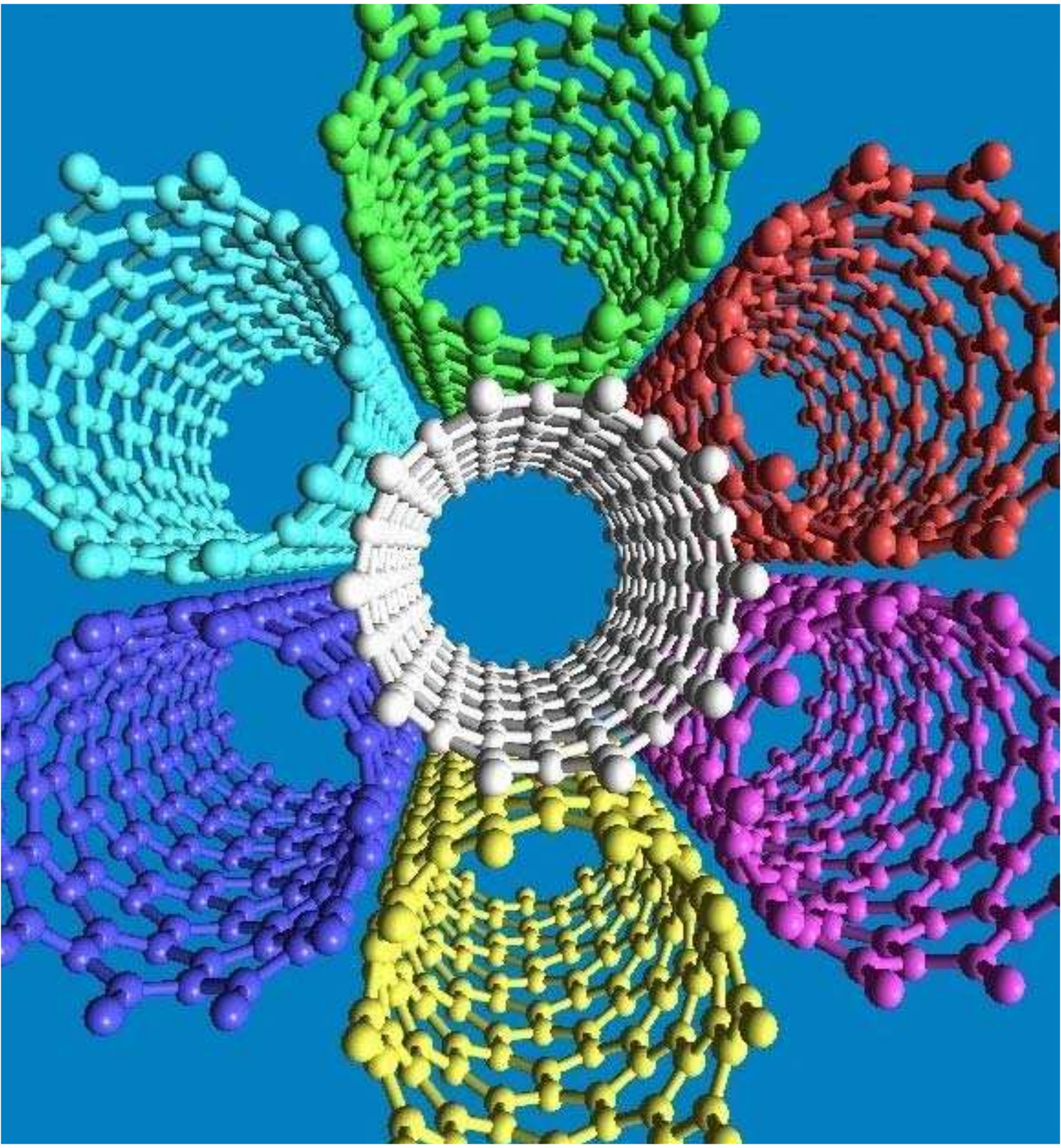}
\caption{(Color on line) Array of nanotubes whose axes are parallel and taken to be aligned in the $z$-direction. In this
paper, we consider a subset forming a bundle, and an infinite periodic array in both $x$ and $y$ directions.} \label{fig1}
\end{figure}

Carbon nanotubes (CNTs) are an allotrope of carbon which have been prepared in various configurations. For
example, a one-atom thick sheet of graphene rolled up into a seamless cylinder with diameter on the order of
a nanometer constitutes a singlewall carbon nanotube (SWNT). The ratio of the length of the resulting
nanostructure  to the diameter exceeds $10^6$. The novel properties of such cylindrical carbon tubules have
made them potentially useful in many applications in nanoscience. These include photonics, electronics
(electrical circuits) and other areas of materials science (to strengthen polymer
materials).\cite{1,2,3,4,5,6} Their extraordinary strength,  unique electrical properties and their  ability
to conduct heat efficiently have generated considerable interest among both experimentalists and
theoreticians.\cite{7,8,9,10,11,12,13,14,15,16,17,18,19,20,21,22,23} Nanotubes may also be produced by
synthesis methods, for instance  multi-wall nanotubes (MWNTs) which consist of multiple layers of graphene
rolled in on themselves to form a cylindrical tube. There are two models which can be used to describe the
structures of multi-wall nanotubes. In the first model, sheets of graphene
 are arranged in concentric cylinders, e.g. a SWNT nanotube  within a larger SWNT
nanotube. In the Parchment model, a single sheet of graphene is rolled in
around itself. The interlayer distance in multi-wall nanotubes is close to the
distance between graphene layers in graphite, approximately $3.3$\ \AA.
Double-wall carbon nanotubes (DWNT) are particularly interesting because they
combine similar morphology and properties compared to SWNT, but improving
significantly their resistance to chemicals.\cite{7,9}
It is abundantly clear that carbon nanotubes are unique nanostructures with remarkable mechanical and
electronic   properties.\cite{1,2,3}  Interest has focused on them as prototypes for a one-dimensional
quantum wire as well as how these one-dimensional properties would be modified when the nanotubes are
combined in a linear array on a two-dimensional (2D) plane or arranged on a three-dimensional lattice. In
Fig.\ \ref{fig1}, we show schematically an array of nanotubes.  We show that the
system of nanotubes would lead to  a coupling between the plasmon modes.


Over the years, there have been a number of experimental measurements \cite{revi1,revi2,revi3,revi4}
and theoretical calculations \cite{revi6,revi7,revi8}
dealing with the effects due to coupling between nanotubes.  In the work of Kociak, et al. \cite{revi1}
employing electron energy loss spectroscopy (EELS),
it   was demonstrated that due to the strong intra-tube and inter-tube Coulomb coupling,
there is an in-plane mode and an out-of-plane mode for an
  array of nanotubes arranged on a two-dimensional surface. This is in agreement
with the work of Gumbs and  A\v{\i}zin \cite{7} who derived the dispersion equation for
nanotubes whose axes are aligned on a planar surface (see also the paper by
Shyu and Lin \cite{10}. St\'{e}phan, et al. \cite{revi2} reported on a detailed experimental
investigation the plasmon excitations of multi-wall carbon nanotubes using EELS.
The results in Ref. \cite{revi2}  compared  the plasmon modes for MWNTs with
those measured for SWNTs.  In interpreting their data \cite{revi2}, a continuum dielectric model
was adopted in which the bulk response function for the material making up the nano-particles
was used. The EELS experiments yielded high energy excitations with energies in the
5-15 eV range. Lower energy plasmon excitations ($\sim 0.5$ eV) were measured
by Murakami, et al. \cite{revi3} in
aligned SWNTs on a planar surface forming a thin film by means of optical absorption techniques,
thereby demonstrating the anisotropy of the modes \cite{7}. Films of thickness 1000 $\AA$
consisting of SWNs with mean diameter $4$ nm have been  probed using EELS
and the plasmon mode dispersion was obtained \cite{revi4}. This means that there are
about twenty-five SWNs stacked on top of each other. Within the limits of the
resolution for the momentum and energy transfer for the EELS experiments
reported in Ref.\ \cite{revi4}. (See also the paper by Thess, et al. \cite{revi5}.)
The observed plasmons are in the energy range of 5-7  eV and wave number $\sim 0.15\ \AA^{-1}$
and are believed to be excitations of the $\pi$-electrons which are formed by
the $2p_z$  orbitals \cite{19}. For metallic and narrow-gap semiconducting SWNTs, the
plasmon excitation energies are of low frequency with energies in the range
$< 1$ eV \cite{19}.  These collective  excitations  are due principally by exciting the
charge carriers in   the low-energy bands near the Fermi level.  With higher
spectroscopic resolution, these low-frequency plasmon excitations should be observed, as we
demonstrate in this paper with the use of a simple model which is valid for this range
of frequencies.  The results for the low-frequency
plasmon excitations in the tight-binding approximation
agree with the electron gas model for SWNTs and DWNTs (see Refs. \cite{19} and \cite{revi7}
as well as references therein) and this is the reason we use the present model for an array.

The electron gas model for  metallic carbon nanotube bundle was
presented by Lin, et al. \cite{revi6} following the reported experiments
of  Thess, et al. \cite{revi5}.  Their formalism for calculating the dispersion
equation for this three-dimensional electron system was based on the RPA.
However,  since  the plasmon excitations cannot be categorized as intrasubband and intersubband plasmons
 as for a single tubule, it is crucial to include the coupling between subbands with different
angular momentum quantum numbers. It is the purpose
of this paper to calculate the  dispersion relation of the low-frequency
plasmons for doped \cite{revi9} nanotubes in which the
charge carriers  are introduced onto the graphene tubules by means of intercalation,
as can be done in carbon fibers or $\rm C_{60}$ \cite{revi10}.
 Also, we investigate  the way in which
 this dispersion relation is affected by their geometrical arrangement,
such as their separation and angular configuration.
 This effect may be analyzed through the behavior of the collective plasma excitations. We do so by calculating
 numerically the plasma dispersion as a function of the separation
between the tubules. We demonstrate that for a bulk lattice the translational
symmetry of the lattice is maintained in the plasmon spectrum, and
the plasmon energies have a periodic dependence on the transverse wave vector
components.

    We first calculate the plasmon dispersion equation for an isosceles right triangle of
nanotubes as a simple example of a bundle of nanotubes on a substrate with their axes aligned parallel to the
$z$ axis.\cite{24,25,26,27,28,29}  The weak van der Waals interaction which holds them together is neglected
in our calculations. The carbon nanotubes are oriented perpendicular to the substrate. With one of the axes
of the tubules on the $z$ axis, the other tubules have their axes at $x=a_x$ on the $x$ axis and $y=a_y$ on
the $y$ axis. In the limit $a_x \to\infty$ or $a_y \to\infty$, our dispersion equation reduces to that for a
pair of multishell nanotubes, as obtained previously.\cite{23} We examine how the Coulomb interaction effects
on the plasmon spectrum depend on the axial separation. In order to consider the role played by the
periodicity of a lattice on the collective modes, we consider an array of nanotubes, with their axes parallel
to the $z$-axis and equally spaced by distance $a_x$ in the $x$-direction and $a_y$ in the $y$-direction.
Each nanotube may consist of $M\geq1$ co-axial cylindrical tubules. We assume that there are no electrons
tunneling between the tubules in each nanotube and between the nanotubes. The plasmons are determined by the
angular momentum quantum number $m$ corresponding to transitions within a subband ($m=0$) or between
different subbands ($m\neq0$) as well as the wave vector $q_z$ along the axis of the nanotube and the
transverse wave vector components $q_x$ and $q_y$.

    Even in the absence of tunneling, for finite $a_x$ and $a_y$,
the plasmon spectrum for one  nanotube is modified by the Coulomb interaction between the nanotubes.\cite{7}
Furthermore, the inter-tubule Coulomb interaction causes the angular momentum not to be conserved in a
nanotube array or bundle, and modes with different $m$ are coupled to one another. In  particular, the
degeneracy of the modes with angular momentum quantum number $m$ and $-m$ is lifted by the Coulomb
interaction.

      The numerical calculations we present were carried out
to analyze the way in which mode-coupling modifies the plasmon spectrum.  For
the bundle consisting of three nanotubes whose axes are at the
vertices of a triangle, we calculate the dispersion relation
as a function of the transverse and longitudinal components of
the wave vector. For the array of nanotubes on a 2D lattice, we
show that the symmetry of
the lattice is maintained in the plasmon spectrum, and the plasmon
excitations depend on the wave vector $q_x$ and $q_y$ with
period $\displaystyle{\frac{2\pi}{a_x}}$ and $\displaystyle{\frac{2\pi}{a_y}}$
in the $x$ and $y$-directions, respectively.
Numerical results are presented for the array of nanotubes on a 2D lattice
with $M=1$. We obtain data for the plasmon excitation energies as functions of
$q_z,\ q_x$ and the lattice period $a_x$.  Different plasmon modes associated with the
intrasubband and intersubband electron transitions in the tubules
are considered and compared with the results for a single nanotube
and a linear array on a 2D planar sheet.

The outline of the rest of this paper is as follows. In Sec. \ref{sec2},
we derive the dispersion relation for plasmons in a bundle of nanotubes
 In Sec. \ref{sec3}, we present and discuss the plasma dispersion equation
for a periodic array of nanotubes consisting of an arbitrary number of concentric tubules. In Sec.
\ref{sec4}, we  present numerical results for the plasmon excitation energies and discuss their properties as
functions of their wave vectors and the separation of the nanotubes. We give a summary in Sec. \ref{sec5}.

\section{Theoretical Formalism for a Bundle of Tubules}
\label{sec2}

We first consider a simple model of bulk quantities of nanotubes. For this
we have a coupled triad of infinitesimally thin nanotubules, with
their axes aligned in the z-direction.  The nanotubes are predicted
to be semiconducting or metallic depending on the chirality of the tubules.
The axis of one of the tubules
of radius  $R_1$ is located at the origin at  $x=0$ and the remaining
two tubules each of radius  $R_2$ and  $R_3$  are located at a distance
of  $x=a_x$ and $y=a_y$   from the origin on the $x$ and $y$-axis,
respectively.  We shall impose the condition that  $a_x>R_1+R_2$
and  $a_y>R_1+R_3$. There is no tunneling between the tubules so
that the eigenfunctions of an electron on the $j^{th}$ tubule ($j=1,2$)
on the $x$-axis or $y$-axis , with axial wave vector   and angular
momentum quantum number  are given by

\begin{eqnarray}
|\nu, j\rangle_x&=& \frac{e^{ik_zz}}{\sqrt{L_z}}\Psi_{l,j}\left( \bm{\rho}-(j-1)a_x \hat{{\bf e}}_x \right)\ ,
\nonumber\\
|\nu, j\rangle_y&=& \frac{e^{ik_zz}}{\sqrt{L_z}}\Psi_{l,j}\left( \bm{\rho}-(i-1)a_y \hat{{\bf e}}_y \right)\ ,
\end{eqnarray}
with $\nu=\{k_z,l\}$ and

\begin{equation}
\Psi_{l,j}(\bm {\rho})=\frac{e^{il\phi}}{\sqrt{2\pi}}\frac{1}{\sqrt{R_j}}\Phi_j\left( \rho \right) \ ,
\hspace{.5cm}\Phi_j^2(\rho)=\delta(\rho-R_j)\ .
\end{equation}
The  energy   eigenvalues are

\begin{equation}
\epsilon_{\nu,j}=\frac{\hbar^2k_z^2}{2m^\ast}+\frac{\hbar^2l^2}{2m^\ast R_j^2}\ .
\end{equation}
 The plasmon dispersion equation can be obtained by solving the density matrix equation
$\displaystyle{i\hbar\frac{d\hat\rho}{dt}=[\hat H,\hat\rho }]$ where
$\hat H=\hat H_0-e\phi_{\rm ext}$ and $\hat\rho=\hat\rho_0+\delta\hat\rho$
with $\langle\nu j|\hat H_0|\nu^\prime j^\prime\rangle=
\epsilon_{\nu,j}\delta_{\nu\nu^\prime}\delta_{jj^\prime}$ and
$\langle\nu j|\hat \rho_0|\nu^\prime j^\prime\rangle=
f_0(\epsilon_{\nu,j})\delta_{\nu\nu^\prime}\delta_{jj^\prime}$,
where $f_0(\epsilon_{\nu,j})$   is the Fermi-Dirac distribution function.
In the lowest order of perturbation theory, we obtain

\begin{equation}
\langle\nu j|\delta\hat \rho_0|\nu^\prime j^\prime\rangle=2e
\frac{f_0(\epsilon_{\nu,j})-f_0(\epsilon_{\nu,j})}
{\hbar\omega-\epsilon_{\nu,j}+\epsilon_{\nu,j}}
\langle\nu j|\phi_{\rm tot}({\bf r},\omega)|\nu^\prime j^\prime\rangle \ .
\end{equation}
Here, $\phi_{\rm tot}({\bf r},\omega)=\phi_{\rm ind}({\bf r},\omega)+
\phi_{\rm ext}({\bf r},\omega)$ is the sum of the external and induced
potentials with $\phi_{\rm ind}({\bf r},\omega)$ a solution
of Poisson's equation

\begin{equation}
\nabla^2\phi_{\rm ind}({\bf r},\omega)=\frac{4\pi e}{\epsilon_s} \delta n_{\rm ind}({\bf r},\omega)\ .
\end{equation}

Now, we obtain the following  matrix elements for the tubules of radii $R_1,\ R_2$ and $R_3$, i.e.,

\begin{eqnarray}
\langle \nu,j\left|e^{i{\bf q}\cdot{\bf r}} \right| \nu^\prime , j^\prime\rangle &=&
\delta_{k_z^\prime,k_z-q_z} e^{-im\lambda} i^m J_m\left( q_\perp R_j \right)\delta_{jj^\prime}\ ,
\end{eqnarray}
where $q_\perp=\sqrt{q_x^2+q_y^2}$. This yields

\begin{eqnarray}
\delta n_{\rm ind}({\bf r},\omega)=\frac{2e}{V}
\sum_{j,j^\prime}\sum_{\nu,\nu^\prime}
\frac{f_0(\epsilon_{\nu,j})-f_0(\epsilon_{\nu^\prime,j^\prime})}
{\hbar\omega-\epsilon_{\nu,j}+\epsilon_{\nu^\prime,j^\prime}}
\sum_{{\bf q},{\bf q}^\prime}\langle \nu^\prime,j^\prime\left|e^{-i{\bf q}\cdot{\bf r}} \right| \nu, j\rangle
\phi_{\rm ind}({\bf q},\omega)
\langle \nu,j\left|e^{i{\bf q}^\prime\cdot{\bf r}} \right| \nu^\prime , j^\prime\rangle
\label{new1}
\end{eqnarray}
and

\begin{eqnarray}
& &\delta n_{\rm ind}({\bf q},\omega)=\frac{2e}{V} \sum_{k_z}
\frac{f_0(\epsilon_{k_z,l;1})-f_0(\epsilon_{k_z-q_z,l-m;1})}
{\hbar\omega-\epsilon_{k_z,l;1}+\epsilon_{k_z-q_z,l-m;1}} e^{-im\vartheta_2}J_m(q_\perp R_1)
\nonumber\\
&\times& \sum_{q_x^\prime,q_y^\prime}\phi_{\rm ind}(q_x^\prime,q_y^\prime,q_z;\omega)
J_m(q_\perp^\prime R_1)
\left(\frac{q_x^\prime-iq_y^\prime}{q_\perp^\prime}  \right)^m
\nonumber\\
&+&
\frac{2e}{V}
\sum_{k_z}
\frac{f_0(\epsilon_{k_z,l;2})-f_0(\epsilon_{k_z-q_z,l-m;2})}
{\hbar\omega-\epsilon_{k_z,l;2}+\epsilon_{k_z-q_z,l-m;2}}
F_1(m,q_\perp^\prime,\vartheta_2; R_2)
\nonumber\\
&\times& \sum_{q_x^\prime,q_y^\prime}\phi_{\rm ind}(q_x^\prime,q_y^\prime,q_z;\omega)
 e^{-iq_x^\prime a_x}F_1(m,q_\perp^\prime,\vartheta_2; R_2)
\nonumber\\
&+&
\frac{2e}{V}
\sum_{k_z}
\frac{f_0(\epsilon_{k_z,l;2})-f_0(\epsilon_{k_z-q_z,l-m;2})}
{\hbar\omega-\epsilon_{k_z,l;2}+\epsilon_{k_z-q_z,l-m;2}}
e^{-iq_ya_y} F_2^\ast (m,q_\perp,\vartheta_3; R_3)
\nonumber\\
&\times& \sum_{q_x^\prime,q_y^\prime}\phi_{\rm ind}(q_x^\prime,q_y^\prime,q_z;\omega) e^{iq_y^\prime
a_y}F_2(m,q_\perp^\prime,\vartheta_3; R_3)\ ,
\end{eqnarray}
where $\vartheta_2$ is the angle between the $x$-axis and a line drawn from the
center of the cylinder at the origin to a point on circumference of the
cylinder with its center at $x=a_x$. We also define $\vartheta_3$ as the angle
between the $y$-axis and a line drawn from the center of the cylinder at the
origin to a point on the circumference of the cylinder located on the $y$-axis
at $y=a_y$. In addition, we have introduced the following notation:

\begin{eqnarray}
F_1(m,q_\perp,\vartheta_2; R_2)&=& e^{im\vartheta_2}J_m(q_\perp R_2)
\nonumber\\
F_2(m,q_\perp,\vartheta_3; R_3)&=& e^{im\vartheta_3}J_m(q_\perp R_3)\ .
\end{eqnarray}

The potential  $\phi_{\rm ind}({\bf q},\omega)$ in Eq. (\ref{new1}) can
be rewritten  in terms of   $\delta n_{\rm ind}({\bf q},\omega)$. This gives

\begin{eqnarray}
\delta n_{\rm ind}({\bf q},\omega)&=&-\ \frac{2e^2}{\varepsilon_s}\chi_{1,m}(q_z,\omega)
J_m(q_\perp R_1)U_{1,m}(q_z )\left(\frac{q_x+iq_y}{q_\perp}  \right)^m
\nonumber\\
&-&  \frac{2e^2}{\varepsilon_s}\chi_{2,m}(q_z,\omega) e^{-iq_xa_x}
F_1(m,q_\perp,\vartheta_2; R_2)U_{2,m}(q_z )
\nonumber\\
&-&  \frac{2e^2}{\varepsilon_s}\chi_{3,m}(q_z,\omega) e^{-iq_ya_y} F_2(m,q_\perp,\vartheta_3; R_3)U_{3,m}(q_z
) \ ,
\end{eqnarray}
with

\begin{eqnarray}
U_{1,m}(q_z )&=& \frac{1}{L_xL_y}\sum_{q_x,q_y}
\frac{\delta n_{\rm ind}(q_x,q_y,q_z)}{q_x^2+q_y^2+q_z^2}
J_m(q_\perp R_1)
\left( \frac{q_x-iq_y}{q_\perp} \right)^m
\nonumber\\
U_{2,m}(q_z )&=& \frac{1}{L_xL_y}\sum_{q_x,q_y}
\frac{\delta n_{\rm ind}(q_x,q_y,q_z)}{q_x^2+q_y^2+q_z^2}
e^{iq_xa_x}
F_1(m,q_\perp,\vartheta_2; R_2)
\nonumber\\
U_{3,m}(q_z )&=& \frac{1}{L_xL_y}\sum_{q_x,q_y}
\frac{\delta n_{\rm ind}(q_x,q_y,q_z)}{q_x^2+q_y^2+q_z^2}
e^{iq_ya_y}
F_2(m,q_\perp,\vartheta_3; R_3)
\end{eqnarray}
and

\begin{equation}
\chi_{j,m}(q_z,\omega)= 2\sum_{l=-\infty}^\infty \int_{-\infty}^\infty dk_z\
\frac{f_0(\epsilon_{k_z,l;j})-f_0(\epsilon_{k_z-q_z,l-m;j})}
{\hbar\omega-\epsilon_{k_z,l;j}+\epsilon_{k_z-q_z,l-m;j}}\ .
\end{equation}
After substituting, we obtain

\begin{eqnarray}
& & U_{1,m}(q_z )= -\frac{2e^2}{\varepsilon_s}\chi_{1,m}(q_z,\omega)
U_{1,m}(q_z )\frac{1}{L_xL_y}\sum_{q_x^\prime,q_y^\prime}
\frac{J_{m^\prime}(q_\perp^\prime R_1) J_{m}(q_\perp^\prime R_1)}
{q_x^{\prime\ 2}+q_y^{\prime\ 2}+q_z^2}
\left( \frac{q_x^\prime+iq_y^\prime}{q_\perp^\prime}
 \right)^{m^\prime-m}
\nonumber\\
&-& \frac{2e^2}{\varepsilon_s}\chi_{2,m}(q_z,\omega)
U_{2,m^\prime}(q_z )\frac{1}{L_xL_y}\sum_{q_x^\prime,q_y^\prime}
e^{-iq_x^\prime a_x}
\frac{F_1^\ast(m^\prime,q_\perp^\prime,\vartheta_2, R_2) J_{m}(q_\perp^\prime R_1)}
{q_x^{\prime\ 2}+  q_z^2}
\nonumber\\
&\times& \left( \frac{q_x^\prime+iq_y^\prime}{q_\perp^\prime}
 \right)^{-m}
\nonumber\\
&-& \frac{2e^2}{\varepsilon_s}\chi_{3,m^\prime}(q_z,\omega)
U_{3,m^\prime}(q_z )\frac{1}{L_xL_y}\sum_{q_x^\prime,q_y^\prime}
e^{-iq_y^\prime a_y}
\frac{F_2(m^\prime,q_\perp^\prime,\vartheta_3, R_3) J_{m}(q_\perp^\prime R_1)}
{q_x^{\prime\ 2}+  q_z^2}
\left( \frac{q_x^\prime+iq_y^\prime}{q_\perp^\prime}
 \right)^{-m} \label{5.1.193}
\end{eqnarray}

\begin{eqnarray}
& & U_{2,m}(q_z )= -\frac{2e^2}{\varepsilon_s}\chi_{1,m}(q_z,\omega)
U_{1,m}(q_z )\frac{1}{L_xL_y}\sum_{q_x^\prime,q_y^\prime} e^{iq_x^\prime a_x}
\frac{F_1(m,q_\perp^\prime,\vartheta_2; R_2) J_{m^\prime}(q_\perp^\prime R_1)}
{q_\perp^{\prime\ 2}+q_z^2}
\nonumber\\
 &\times&\left( \frac{q_x^\prime+iq_y^\prime}{q_\perp^\prime}
 \right)^{m^\prime}
\nonumber\\
 &-& \frac{2e^2}{\varepsilon_s}\chi_{2,m^\prime}(q_z,\omega)
U_{2,m^\prime}(q_z )\frac{1}{L_xL_y}\sum_{q_x^\prime,q_y^\prime}
\frac{F_1^\ast(m^\prime,q_\perp^\prime,\vartheta_2, R_2) F_1(m,q_\perp^\prime,\vartheta_2, R_2)}
{q_\perp^{\prime\ 2}+  q_z^2}e^{-i(q_x-q_x^\prime)a_x}
\nonumber\\
&-& \frac{2e^2}{\varepsilon_s}\chi_{3,m^\prime}(q_z,\omega)
U_{3,m^\prime}(q_z )\frac{1}{L_xL_y}\sum_{q_x^\prime,q_y^\prime}
\frac{F_2^\ast(m^\prime,q_\perp^\prime,\vartheta_3, R_3)
F_1(m,q_\perp^\prime,\vartheta_2, R_2)}
{q_\perp^{\prime\ 2}+  q_z^2}
\nonumber\\
&\times&   e^{-iq_y a_y} e^{iq_x^\prime a_x} \label{5.1.194}
\end{eqnarray}

\begin{eqnarray}
& & U_{3,m}(q_z )= -\frac{2e^2}{\varepsilon_s}\chi_{1,m^\prime}(q_z,\omega)
U_{1,m^\prime}(q_z )\frac{1}{L_xL_y}\sum_{q_x^\prime,q_y^\prime} e^{iq_y^\prime a_y}
\frac{F_2(m,q_\perp^\prime,\vartheta_3; R_3) J_{m^\prime}(q_\perp^\prime R_1)}
{q_\perp^{\prime\ 2}+q_z^2}
\nonumber\\
&\times& \left( \frac{q_x^\prime+iq_y^\prime}{q_\perp^\prime}
 \right)^{m^\prime}
\nonumber\\
 &-& \frac{2e^2}{\varepsilon_s}\chi_{2,m^\prime}(q_z,\omega)
U_{2,m^\prime}(q_z )\frac{1}{L_xL_y}\sum_{q_x^\prime,q_y^\prime}
\frac{F_1^\ast(m^\prime,q_\perp^\prime,\vartheta_2, R_2)
F_2(m,q_\perp^\prime,\vartheta_3, R_3)}
{q_\perp^{\prime\ 2}+  q_z^2}
 e^{-iq_y a_y}  e^{iq_y^\prime a_y}
\nonumber\\
&-& \frac{2e^2}{\varepsilon_s}\chi_{3,m^\prime}(q_z,\omega) U_{3,m^\prime}(q_z
)\frac{1}{L_xL_y}\sum_{q_x^\prime,q_y^\prime} \frac{F_2^\ast(m^\prime,q_\perp^\prime,\vartheta_3, R_3)
F_2(m,q_\perp^\prime,\vartheta_3, R_3)} {q_\perp^{\prime\ 2}+  q_z^2} e^{-iq_y^\prime a_y} e^{iq_y a_y}
\nonumber\\
&& \label{5.1.195}
\end{eqnarray}
We note that the set of equations in Eqs. (\ref{5.1.193}) - (\ref{5.1.195}) form a a simultaneous system in
the variables $U_{1,m}(q_z ),\ U_{2,m}(q_z )$ and $U_{3,m}(q_z )$, with $m=0,\pm 1,\pm 2,\cdots$.  These
equations may be simplified if we now introduce the following quantities below, i.e.,  if we set

\begin{eqnarray}
A_{mm^\prime}&=& \frac{2e^2}{\varepsilon_s}\chi_{1,m^\prime}(q_z,\omega)
\frac{1}{L_xL_y}\sum_{q_x^\prime,q_y^\prime} \frac{J_{m^\prime}(q_\perp^\prime R_1) J_{m}(q_\perp^\prime
R_1)} {q_\perp^{\prime\ 2}+q_z^2} \left( \frac{q_x^\prime+iq_y^\prime}{q_\perp^\prime}
 \right)^{m^\prime-m}
\nonumber\\
B_{mm^\prime}&=& \frac{2e^2}{\varepsilon_s}\chi_{2,m^\prime}(q_z,\omega)
\frac{1}{L_xL_y}\sum_{q_x^\prime,q_y^\prime}
e^{-iq_x^\prime a_x}
\frac{F_1(m^\prime ,q_\perp^\prime,\vartheta_2, R_2) J_{m}(q_\perp^\prime R_1)}
{q_\perp^{\prime\ 2}+q_z^2}
\left( \frac{q_x^\prime+iq_y^\prime}{q_\perp^\prime}
 \right)^{-m}
\nonumber\\
C_{mm^\prime}&=& \frac{2e^2}{\varepsilon_s}\chi_{3,m^\prime}(q_z,\omega)
\frac{1}{L_xL_y}\sum_{q_x^\prime,q_y^\prime}
e^{-iq_y^\prime a_y}
\frac{F_2(m^\prime ,q_\perp^\prime,\vartheta_3, R_3) J_{m^\prime}(q_\perp^\prime R_1)}
{q_\perp^{\prime\ 2}+q_z^2}
\left( \frac{q_x^\prime+iq_y^\prime}{q_\perp^\prime}
 \right)^{-m}
\nonumber\\
D_{mm^\prime}&=& \frac{2e^2}{\varepsilon_s}\chi_{1,m^\prime}(q_z,\omega)
\frac{1}{L_xL_y}\sum_{q_x^\prime,q_y^\prime}
e^{iq_x^\prime a_x}
\frac{F_1(m ,q_\perp^\prime,\vartheta_2, R_2) J_{m^\prime}(q_\perp^\prime R_1)}
{q_\perp^{\prime\ 2}+q_z^2}
\left( \frac{q_x^\prime+iq_y^\prime}{q_\perp^\prime}
 \right)^{m^\prime}
\nonumber\\
E_{mm^\prime}&=& \frac{2e^2}{\varepsilon_s}\chi_{2,m^\prime}(q_z,\omega)
\frac{1}{L_xL_y}\sum_{q_x^\prime,q_y^\prime}
\frac{F_1^\ast(m^\prime ,q_\perp^\prime,\vartheta_2, R_2)
F_1(m ,q_\perp^\prime,\vartheta_2, R_2)}
{q_\perp^{\prime\ 2}+q_z^2}
e^{-i(q_x -q_x^\prime)a_x}
\nonumber\\
F_{mm^\prime}&=& \frac{2e^2}{\varepsilon_s}\chi_{3,m^\prime}(q_z,\omega)
\frac{1}{L_xL_y}\sum_{q_x^\prime,q_y^\prime}
\frac{F_2^\ast(m^\prime ,q_\perp^\prime,\vartheta_3, R_3)
F_1(m ,q_\perp^\prime,\vartheta_2, R_2)}
{q_\perp^{\prime\ 2}+q_z^2}
e^{-iq_y a_y} e^{iq_x^\prime a_x}
\nonumber\\
G_{mm^\prime}&=& \frac{2e^2}{\varepsilon_s}\chi_{1,m^\prime}(q_z,\omega)
\frac{1}{L_xL_y}\sum_{q_x^\prime,q_y^\prime}
e^{iq_y^\prime a_y}
\frac{F_2(m ,q_\perp^\prime,\vartheta_3, R_3) J_{m^\prime}(q_\perp^\prime R_1)}
{q_\perp^{\prime\ 2}+q_z^2}
\left( \frac{q_x^\prime+iq_y^\prime}{q_\perp^\prime}
 \right)^{m^\prime}
\nonumber\\
H_{mm^\prime}&=& \frac{2e^2}{\varepsilon_s}\chi_{2,m^\prime}(q_z,\omega)
\frac{1}{L_xL_y}\sum_{q_x^\prime,q_y^\prime}
\frac{F_1^\ast(m,q_\perp^\prime,\vartheta_2, R_2)
F_2(m ,q_\perp^\prime,\vartheta_3, R_3)}
{q_\perp^{\prime\ 2}+q_z^2}
e^{-iq_x a_x} e^{iq_y^\prime a_y}
\nonumber\\
I_{mm^\prime}&=& \frac{2e^2}{\varepsilon_s}\chi_{3,m^\prime}(q_z,\omega)
\frac{1}{L_xL_y}\sum_{q_x^\prime,q_y^\prime} \frac{F_2^\ast(m^\prime,q_\perp^\prime,\vartheta_3, R_3) F_2(m
,q_\perp^\prime,\vartheta_3, R_3)} {q_\perp^{\prime\ 2}+q_z^2} e^{-iq_y a_y} e^{iq_y^\prime a_y}\ .
\label{5.1.196}
\end{eqnarray}
We note that $F_{mm}$ and $H_{mm}$ are the only matrix elements containing a product of the two factors
$\exp(iq_\perp a_x\cos\vartheta)$ and $\exp(-iq_\perp a_y\sin\vartheta)$.
 The Jacobi-Auger formula along with a standard formula for
$\exp(iz \cos\vartheta)$ and $\exp(\mp i\sin\vartheta)$ produces the above  obtained results.

Back-substituting the results for the matrix elements in Eq. (\ref{5.1.196}) into Eqs. (\ref{5.1.193}),
(\ref{5.1.194}) and (\ref{5.1.195}), respectively, yields the following compact system of simultaneous linear
equations

\begin{eqnarray}
& &U_{1,m}+\sum_{m^\prime}A_{mm^\prime}U_{1,m^\prime}+
\sum_{m^\prime}B_{mm^\prime}U_{2,m^\prime}
+\sum_{m^\prime}C_{mm^\prime}U_{3,m^\prime}=0
\nonumber\\
& &U_{2,m}+\sum_{m^\prime}D_{mm^\prime}U_{1,m^\prime}+
\sum_{m^\prime}E_{mm^\prime}U_{2,m^\prime}
+\sum_{m^\prime}F_{mm^\prime}U_{3,m^\prime}=0
\nonumber\\
& &U_{3,m}+\sum_{m^\prime}G_{mm^\prime}U_{1,m^\prime}+ \sum_{m^\prime}H_{mm^\prime}U_{2,m^\prime}
+\sum_{m^\prime}I_{mm^\prime}U_{3,m^\prime}=0\ . \label{mode-coupling}
\end{eqnarray}
The mode  coupling arising from the Coulomb interaction is clearly seen through the finite values of the
coefficients when $m^\prime\neq m$ in Eq.\ (\ref{mode-coupling}). The dimension of the associated coefficient
matrix above is dependent on the type of transitions being investigated. Therefore, to obtain a matrix of
modest dimensions, let us consider only     intrasubband transitions (i.e.,  $m=0$).   For such transitions, we
have

\begin{equation}
\tensor{\epsilon}(\omega,q_x,q_y,q_z) \left(\matrix{U_{1,0}(q_z)\cr U_{2,0}(q_z)\cr U_{3,0}(q_z)\cr} \right)
\equiv \left(\matrix{1+A_{0,0}&B_{0,0}&C_{0,0}\cr D_{0,0}&1+E_{0,0}&F_{0,0}\cr G_{0,0}&H_{0,0}&1+I_{0,0}\cr}
\right) \left(\matrix{U_{1,0}(q_z)\cr U_{2,0}(q_z)\cr U_{3,0}(q_z)\cr} \right) =\left(\matrix{0\cr
 0\cr
 0\cr}
\right)\ . \label{5.1.211}
\end{equation}
For non trivial solutions of Eq. (\ref{5.1.211}) to exist, one requires that
 the determinant of the coefficient matrix $\tensor{\epsilon}(q_x,q_y,q_z)$
vanish.  That is $Det\ \tensor{\epsilon}(q_x,q_y,q_z)=0$. This is the dispersion formula for the plasmons and
particle-hole mode  intrasubband transitions only.
     Let us further investigate the intrasubband ($m^\prime=m=0$) plasma dispersion
equation. In general, an explicit determination of the elements of the matrix yields for $m=m^\prime$ leads
to a simplification of the results in Eq.\ (\ref{5.1.196}) and we have

\begin{eqnarray}
A_{mm}&=& \frac{e^2}{\pi\varepsilon_s}I_{m}(q_z R_1)K_{m}(q_z R_1)
\chi_{1,m}(q_z,\omega)
\nonumber\\
B_{mm}&=& \frac{e^2}{\pi\varepsilon_s}\chi_{2,m}(q_z,\omega)
\int_0^\infty dq_\perp\ q_\perp
\frac{J_{0}(q_\perp a_x)  J_{m}(q_\perp R_1)J_{m}(q_\perp R_2)}
{q_\perp^2+q_z^2}
\nonumber\\
C_{mm}&=& \frac{e^2}{\pi\varepsilon_s}\chi_{3,m}(q_z,\omega)
\int_0^\infty dq_\perp\ q_\perp
\frac{J_{0}(q_\perp a_y)  J_{m}(q_\perp R_1)J_{m}(q_\perp R_3)}
{q_\perp^2+q_z^2}
\nonumber\\
D_{mm}&=& \frac{e^2}{\pi\varepsilon_s}\chi_{1,m}(q_z,\omega)
\int_0^\infty dq_\perp\ q_\perp
\frac{J_{0}(q_\perp a_x)  J_{m}(q_\perp R_1)J_{m}(q_\perp R_2)}
{q_\perp^2+q_z^2}
\nonumber\\
E_{mm}&=& \frac{e^2}{\pi\varepsilon_s}I_{m}(q_z R_2)K_{m}(q_z R_2)
\chi_{2,m}(q_z,\omega)
\nonumber\\
F_{mm}&=& \frac{e^2}{\pi\varepsilon_s}\chi_{3,m}(q_z,\omega)
\int_0^\infty dq_\perp\ q_\perp\sum_{k,k^\prime=-\infty}^\infty
\frac{i^k J_{k}(q_\perp a_x)  J_{k^\prime}(q_\perp a_y) J_{m}(q_\perp R_2)  J_{m}(q_\perp R_3)}
{q_\perp^2+q_z^2}
\nonumber\\
G_{mm}&=& \frac{e^2}{\pi\varepsilon_s}\chi_{1,m}(q_z,\omega)
\int_0^\infty dq_\perp\ q_\perp
\frac{J_{0}(q_\perp a_y)  J_{m}(q_\perp R_1)J_{m}(q_\perp R_3)}
{q_\perp^2+q_z^2}
\nonumber\\
H_{mm}&=& \frac{e^2}{\pi\varepsilon_s}\chi_{2,m}(q_z,\omega)
\int_0^\infty dq_\perp\ q_\perp\sum_{k,k^\prime=-\infty}^\infty
\frac{i^k (-1)^{k^\prime} J_{k}(q_\perp a_x)  J_{k^\prime}(q_\perp a_y) J_{m}(q_\perp R_2)  J_{m}(q_\perp R_3)}
{q_\perp^2+q_z^2}
\nonumber\\
I_{mm}&=& \frac{e^2}{\pi\varepsilon_s}I_{m}(q_z R_3)K_{m}(q_z R_3) \chi_{3,m}(q_z,\omega)\ .
\label{mmelements}
\end{eqnarray}
There is symmetry between the matrix elements appearing in Eq.\ (\ref{mmelements}). For example, the only
difference between $B_{mm}$ in the first row and second column and $D_{mm}$ in the second row and first
column is the susceptibility $\chi_{j,m}(q_z,\omega)$, with $j=1$ or $j=2$.  Because of the symmetry under
the interchange $R_1\leftrightarrow R_2$, the form factor in these two matrix elements is the same. Also, the
matrix element $C_{mm}$ may be obtained from $G_{mm}$ with the replacement $\chi_{3,m}(q_z,\omega)\to
\chi_{1,m}(q_z,\omega)$ and vice versa. For these two matrix elements, there is symmetry  under the
interchange $R_1\leftrightarrow R_3$.  Further simplification of these results may be achieved for
intrasubband excitation ($m=m^\prime=0$) by means of the identity \cite{threeJs}

\begin{eqnarray}
\int_0^\infty dq_\perp \frac{q_\perp}{q_\perp^2+q_z^2}J_0(a q_\perp)J_0(b q_\perp)
J_0(c q_\perp)&=& I_0(q_za) I_0(q_zc)K_0(q_zb)
\nonumber\\
& &{\rm for}\ \ a>b+c; \ \ b,c,q_z>0\ .
\end{eqnarray}
In the limit when $a_x\to\infty$, i.e., when the tubule on the $x$-axis is infinitely far away, the
inter-tubule Coulomb interaction is negligible, and we obtain the following dispersion formula for two
tubules on the $y$-axis, i.e.,

\begin{equation}
{\rm Det}\ \left(\matrix{1+A_{0,0}&0&C_{0,0}\cr 0&1+E_{0,0}&0\cr G_{0,0}&0&1+I_{0,0}\cr} \right)=0\ .
\label{case1}
\end{equation}
In the limit $a_y\to\infty$, the dispersion equation becomes

\begin{equation}
{\rm Det}\
\left(\matrix{1+A_{0,0}&B_{0,0}&0\cr
D_{0,0}&1+E_{0,0}&0\cr
0&0&1+I_{0,0}\cr}
\right)=0\ .
\label{case2}
\end{equation}
Thus, as expected, the dispersion equations (\ref{case1}) and (\ref{case2}) clearly demonstrate
a coupling of the modes in a pair of the nanotubes leaving the third isolated from the other
two in the bundle. We now turn to a consideration of the dispersion equation for plasma
excitations for a 2D array of nanotubes embedded in a background dielectric medium.

\section{Two-Dimensional Array of Nanotubes}
\label{sec3}

     Assuming that there is no tunneling between the tubules, the
single-particle eigenfunctions
 for the 2D periodic array are

\begin{eqnarray}
\psi_{j\nu l}(\bm{\rho},z)&=&\frac{1}{\sqrt{L_zN_xN_y}}e^{ik_zz}\sum_{n_x=-\frac{N_x}{2}}
^{\frac{N_x}{2}}
\sum_{n_y=-\frac{N_y}{2}}^{\frac{N_y}{2}}e^{i(k_xn_xa_x+k_yn_ya_y)}
\Psi_{jk_zl}(\bm{\rho}-(n_xa_x\hat{\bf e}_x+n_ya_y\hat{\bf e}_y))\ ,
\nonumber\\
\Psi_{jl}(\bm{\rho})&=&\frac{1}{\sqrt{2\pi}}e^{il\phi}
\frac{1}{\sqrt{R_j}}\Phi_j(\rho)\ ,
\label{gg1}
\end{eqnarray}
where $j=1,2,\cdots,M$ labels the tubules in the nanotube,
$\nu=\{k_x,k_y,k_z\}$ is a composite index for the electron eigenstates,
$\Psi_{jl}(\bm{\rho})e^{ik_zz}$  is the wave function for an  electron in the
$j$-th tubule, with wave vector $k_z$ in the axial direction and angular
momentum quantum number $l=0,\pm1,\pm2,\cdots$,
$\Phi_i^2(\rho)=\delta(\rho-R_j)$, $k_x=\frac{2\pi }{L_x}n_x$ and
$k_y=\frac{2\pi}{L_y}n_y$ with $n_x=0,\pm1,\pm2,\cdots,\pm\frac{N_x}{2}$ and
$n_y=0,\pm1,\pm2,\cdots,\pm\frac{N_y}{2}$. Here,  $N_x=L_x/a_x$ and
$N_y=L_y/a_y$ are the numbers of nanotubes in the $x$ and $y$ directions in the
array with periodic boundary conditions. Electron motion in the azimuthal
direction around the tubule is quantized and characterized by the angular
momentum quantum number $l$, whereas motion in the axial $z$ direction is free.
Thus, the electron spectrum in each tubule consists of one-dimensional (1D)
subbands with $l$ serving as a subband index. The spectrum does not depend on
$k_x$ and has the form

\begin{equation}
\epsilon_{jk_zl}=\frac{\hbar^2k_z^2}{2m^{\ast}}+\frac{\hbar^2l^2}{2m^{\ast}R_j^2}\ . \label{gg2}
\end{equation}
Consequently, using these results in conjunction with the methods previously employed for the one-dimensional
array of nanotubes, we obtain the dispersion formula for plasma excitations in a 2D array of nanotubes to be

\begin{eqnarray}
& &{\rm Det} \left[\delta_{mm^{\prime}}\delta_{jj^{\prime}}+
\frac{2e^2}{a_xa_y \epsilon_s}\chi_{j^{\prime}m^{\prime}}(q_z,\omega) \right.
\sum_{N_x=-\infty}^{\infty}\sum_{N_y=-\infty}^{\infty}
\nonumber\\
&\times&\left.
\frac{J_{m^{\prime}}\left(R_{j^\prime}\sqrt{\left(q_x+G_N^x\right)^2+\left(q_y+G_N^y\right)^2}
 \right)J_m\left(R_j\sqrt{\left(q_x+G_N^x\right)^2+\left(q_y+G_N^y\right)^2} \right)}
{\left(q_x+G_N^x\right)^2+\left(q_y+G_N^y\right)^2+q_z^2}\right.
\nonumber\\
&\times&\left.
\left(\frac{q_x+G_N^x+i\left(q_y+G_N^y\right)}
{\sqrt{(q_x+G_N^x)^2+\left(q_y+G_N^y\right)^2}}  \right)^{m^\prime-m}\right] =0\ ,
\label{gg16+}
\end{eqnarray}
where $G_N^x=\frac{2\pi N_x}{a_x}$ and $G_N^y=\frac{2\pi N_y}{a_y}$.

The result  in Eq.\ (\ref{gg16+}) shows that the symmetry of the lattice is
maintained in the plasmon spectrum for both the $x$ and $y$ directions.  In
addition, the plasmon excitations depend on the wave vector components $q_x$
and $q_y$ with period $\displaystyle{\frac{2\pi}{a_x}}$ and
$\displaystyle{\frac{2\pi}{a_y}}$.  In the limit, $a_y\to\infty$, we obtain the
following dispersion equation for a linear array of nanotubes on a 2D plane,
i.e.,

\begin{eqnarray}
&Det&\left[\delta_{mm^{\prime}}\delta_{jj^{\prime}}+
\frac{e^2}{\pi a_x\epsilon_s}\chi_{j^{\prime}m^{\prime}}(q_z,\omega)\right.
\nonumber\\
&\times&\left.
\sum_{N_x=-\infty}^{\infty}\int_{-\infty}^{\infty} dq_y\
\frac{J_{m^{\prime}}\left(\sqrt{\left(q_x+G_N^x\right)^2+q_y^2}R_{j^{\prime}}
 \right)J_m\left(\sqrt{\left(q_x+G_N^x\right)^2+q_y^2}R_j \right)}
{\left(q_x+G_N^x\right)^2+q_y^2+q_z^2}\right.
\nonumber\\
&\times&\left.
\left(\frac{q_x+G_N^x+iq_y}{\sqrt{(q_x+G_N^x)^2+q_y^2}}  \right)^{m^\prime-m}\right] =0\ ,
\label{gg16}
\end{eqnarray}
which agrees with the result obtained by Gumbs and  A\v{\i}zin.\cite{7}

\section{Numerical Results and Discussion}
\label{sec4}

 We now turn to numerical calculations for an array of nanotubes as well as a bundle based on the
 formalism we developed in Secs.\ \ref{sec2} and \ref{sec3}. We simulate a metallic graphene tubule in
 a medium with background dielectric constant  $\varepsilon_s=2.4$ by choosing  $ m^\ast=0.25 m_e$
where $m_e$ is the free electron mass and  $E_F=0.6$\ eV. The effective Bohr
radius is $a_B\equiv\hbar^2\epsilon_s/m^{\ast}e^2=1.26$\ \AA. All calculations
were carried out at zero temperature. We included the transitions $m=0,\pm 1$
only in the calculations for plasma excitations in a 2D array. The number of
occupied subbands in each tubule is determined by its radius, the electron
density through the Fermi energy  and electron effective mass.  The occupied
subbands are then included in the sum
 $l=0,\pm 1,\pm2,\cdots,\pm l_{\rm max}$, where $l_{\rm max}$ labels the highest occupied subband at $T=0$ K.
For a singlewall cylindrical nanotube of radius  $11$\ \AA,  there are five
subbands occupied by electrons corresponding to $l=0,\pm 1,\pm2$.
 Lin and Shung \cite{9} used the same values of $\varepsilon_s,\ m^{\ast},\ R$ and $ E_F$
in calculating the plasmon excitation spectrum. It was shown that
there are three quasi-acoustic plasmon branches associated with
intrasubband electron excitations with angular momentum transfer
$m=0$. The reason for this is that the plasmon excitations depend on $|l|$ in this case.
 There are five optical plasmon branches associated with intersubband
electron transitions with angular momentum transfer $m=\pm1$.

\begin{figure}[p]
\includegraphics[width=4.5in,height=5.6in]{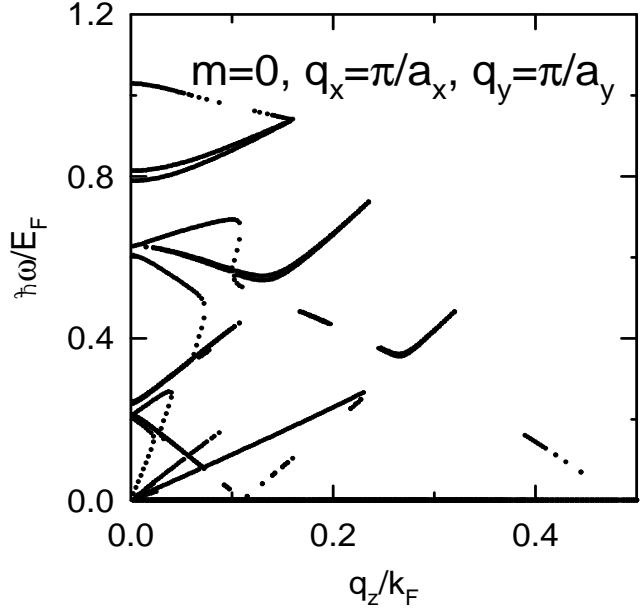}
\caption{Undamped intrasubband ($m=0$) plasmon excitation energy, in units of
the Fermi energy  $E_F$, as a function of $q_z/k_F$. Here, $k_F$ is the Fermi
wave number  in the ground ($l=0$) subband, obtained. The solutions were
obtained  by solving Eq.\ (\ref{gg16+}) for $q_x=\pi/a_x$ and $q_y=\pi/a_y$.
The parameters used in the calculation are $\varepsilon_s=2.4,\ m^{\ast}=0.25
m_e$ where $m_e$ is the bare electron mass and $a_x=a_y=35.0$ {\AA},\ \
$R=11.0$ {\AA},\ \ $E_F=0.6 eV$. } \label{fig2}
\end{figure}

\begin{figure}[p]
\includegraphics[width=4.5in,height=5.6in]{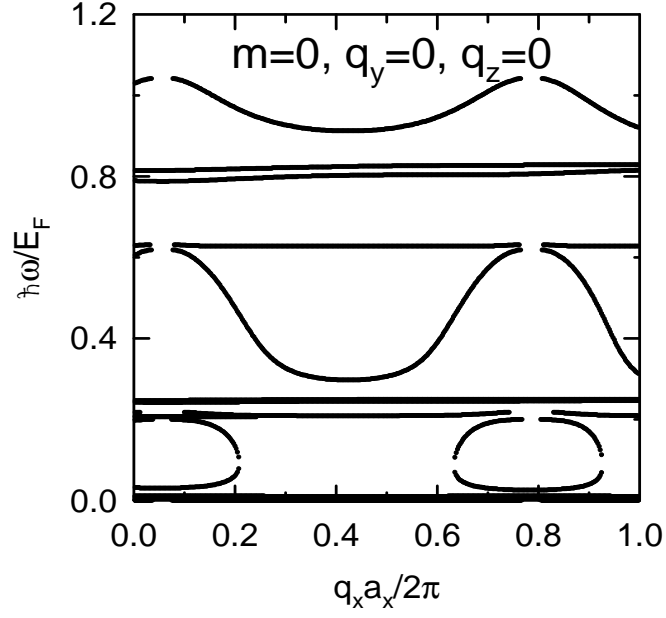}
\caption{Undamped intrasubband ($m=0$) plasmon excitation energy as a function
of the transverse wave vector $q_x$ (in units of $2\pi/a_x$). The solutions are
based on Eq.\ (\ref{gg16+}) for $q_y=0$ and $q_z=0$. All other parameters
employed in obtaining Fig.\ \ref{fig2} are the same.} \label{fig3}
\end{figure}

In Fig.\ \ref{fig2}, we present our results for the dispersion relation of
intrasubband ($m=0$) plasmon excitations of a 2D array of singlewall nanotubes.
The radius of each tubule is $11$ {\AA} and the period of the lattice is $35$\
\AA. The excitation energy is plotted as a function of $q_z$ with $q_x=\pi
/a_x$ and $q_y=\pi/a_y$. Only the plasmon branches which are not Landau damped
and which lie outside the single-particle excitation regions are presented in
Fig.\ \ref{fig2}. For the sake of clarity, we have omitted the boundaries of
the particle-hole continuum in this figure. In Fig.\ \ref{fig3}, we plot the
intrasubband plasmon energies as a function of $q_xa_x/2\pi$ for fixed $q_y$
and $q_z$, using the same lattice parameters as  Fig.\ \ref{fig2}. Only the
plasmon modes which are not Landau damped by the single-particle excitations
are shown in  Fig.\ \ref{fig3}. There are several branches of plasmon
excitations in both Figs.\ \ref{fig2} and    \ref{fig3}. The periodicity of the
lattice is preserved in each of the plasmon branches when plotted as functions
of the transverse wave vector $q_x$.

It is a simple matter to show from Eq.\ (\ref{gg16}) that for a 1D array, when $q_x=0$ the elements of the
determinantal matrix with $m=0,\ m^\prime=\pm 1$ are zero due to antisymmetry of the $q_y$-integrand. This
means that there is decoupling of the intrasubband and intersubband excitations. Similarly, it follows from
Eq.\ (\ref{gg16+}) that when {\em both\/} $q_x=0$ and $q_y=0$, the matrix elements of the determinantal
equation are zero for $m=0$ and $m^\prime=\pm1$ as well. The  main result of these plots is to show that the
Coulomb interaction between the tubules in the 2D array serves to alter substantially the dependence of the
plasmon frequency on $q_z$.  Furthermore, as the separation between  tubules is decreased, the plasmon
frequency is increased due to the role played by the Coulomb interaction between the tubules. Some of the
plasmon branches in Fig.\ \ref{fig3} show a stronger dependence on $q_x$ than others.
 This variation is more pronounced for the high-frequency plasmon mode compared with some of the lower-lying
 frequency ones.

\begin{figure}[p]
\includegraphics[width=4.5in,height=5.6in]{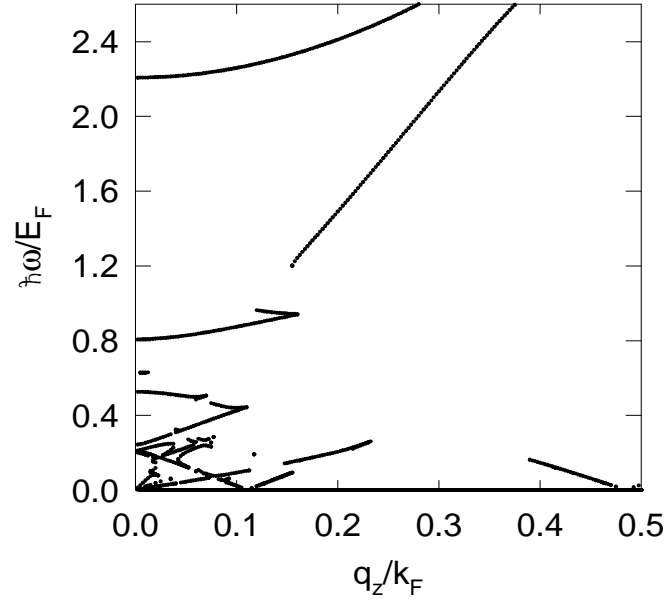}
\caption{The plasmon excitation energy, in units of the Fermi energy  $E_F$, as
a function of $q_z/k_F$, for a 2D array of coaxial tubules of outer and inner
radius $R_1=14$ {\AA} and $R_2=11$ {\AA}, respectively. The lattice constants
in both $x$ and $y$ directions was chosen as $35$ {\AA}. The solutions were
obtained from Eq.\ (\ref{gg16+}), when both $a_x,\ a_y\to\infty$. The
parameters used in the calculation are the same as in Fig. \ref{fig2}.}
\label{fig4}
\end{figure}

 The effect of the Coulomb interaction on the intrasubband plasmon excitation spectrum of
 a 2D array of  nanotubes  consisting of a pair of coaxial tubules is demonstrated in
 Fig.\ \ref{fig4}. The outer radius is $R_1=14.4$\  \AA \ and the inner radius is $R_2=11$\ \AA.
 The period of the lattice was chosen as $a_x=a_y=35$\ \AA\  in both the $x$ and $y$ directions.
 Only those branches which are not Landau damped by  single-particle
 excitations are shown in this figure. The two highest modes originate from the high-energy modes of
 each  tubule. The Coulomb interaction between the coaxial tubules leads to optical branches which
 are not present for a single tubule, all of whose modes are acoustic in nature.

\begin{figure}[p]
\includegraphics[width=4.5in,height=5.6in]{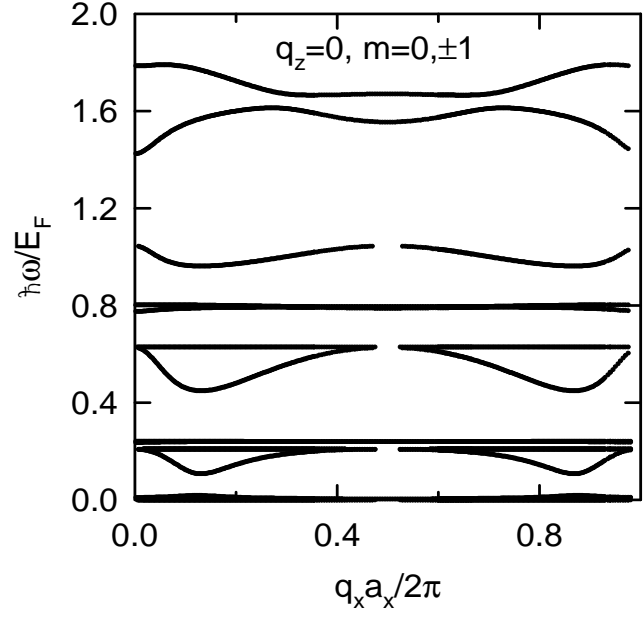}
\caption{For a 1D periodic array, the plasmon excitation energy, in units of
the Fermi energy $E_F$, as a function of $q_x$, in units of $2\pi /a_x$.  Here
$q_z=0$, the radius of each tubule is $11$ {\AA} and the period is $a_x=35$
{\AA}. All other parameters are the
 same as in Fig. \ref{fig2}. }
\label{fig5}
\end{figure}

\begin{figure}[p]
\includegraphics[width=4.5in,height=5.6in]{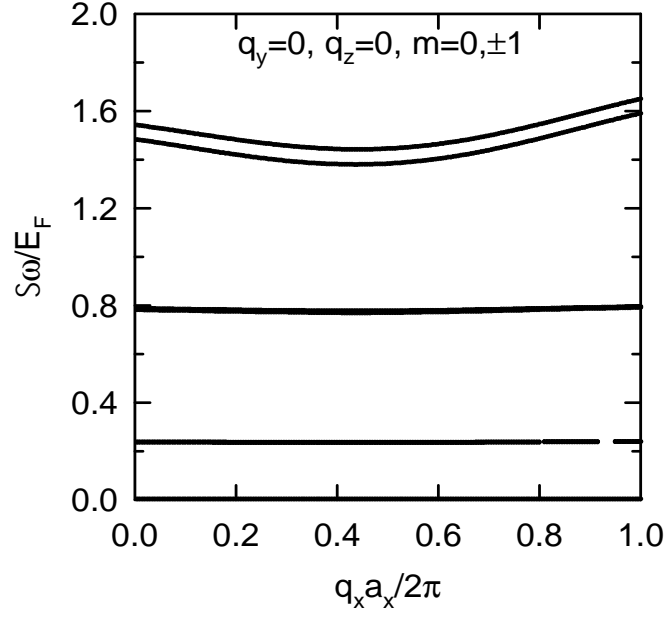}
\caption{For a 2D periodic array, the plasmon excitation energy, in units of the Fermi energy $E_F$, as a
function of $q_x$, in units of $2\pi /a_x$.  Here $q_y=q_z=0$, all other parameters are the
 same as in Fig. \ref{fig2}. }
\label{fig6}
\end{figure}

Plasmon excitations arising from intersubband  electron transitions for $m=\pm
1$ are shown in Figs.\ \ref{fig5} and \ref{fig6} for the 1D and 2D periodic
arrays of single-wall nanotubes. In both figures, only undamped plasmon modes
are shown and the particle-hole continuum is omitted. The highest modes have a
stronger dependence on the transverse wave vector $q_x$ than the lower modes.
This is more so for the 1D array than for the 2D array.  The main difference
between these intersubband  modes in a single tubule compared with a linear 1D
array of tubules   and when they are arranged in a  2D lattice may be explained
in the following way. In a single tubule, plasmon modes with $m=+1$ and $-1$
are degenerate \cite{9}. The periodic lattice of tubules in either one
direction or in two directions removes the axial symmetry of the structure. The
Coulomb interaction between the tubules lifts the degeneracy and splits each
single tubule intersubband plasmon mode into two branches increasing the total
number of modes to ten for both the 1D and 2D lattices. The Coulomb interaction
seems to have a larger effect on the lowest modes for the 1D lattice.
Furthermore, this splitting is larger for the high-frequency modes compared
with the low-frequency ones. This is demonstrated in Figs.\ \ref{fig5} and
\ref{fig6}. As a matter of fact,  the splitting of the low-frequency modes is
too small to be resolved on the scale used in the figures.

\begin{figure}[p]
\centering
\subfloat{%
\resizebox{4.5in}{3.5in}{\includegraphics{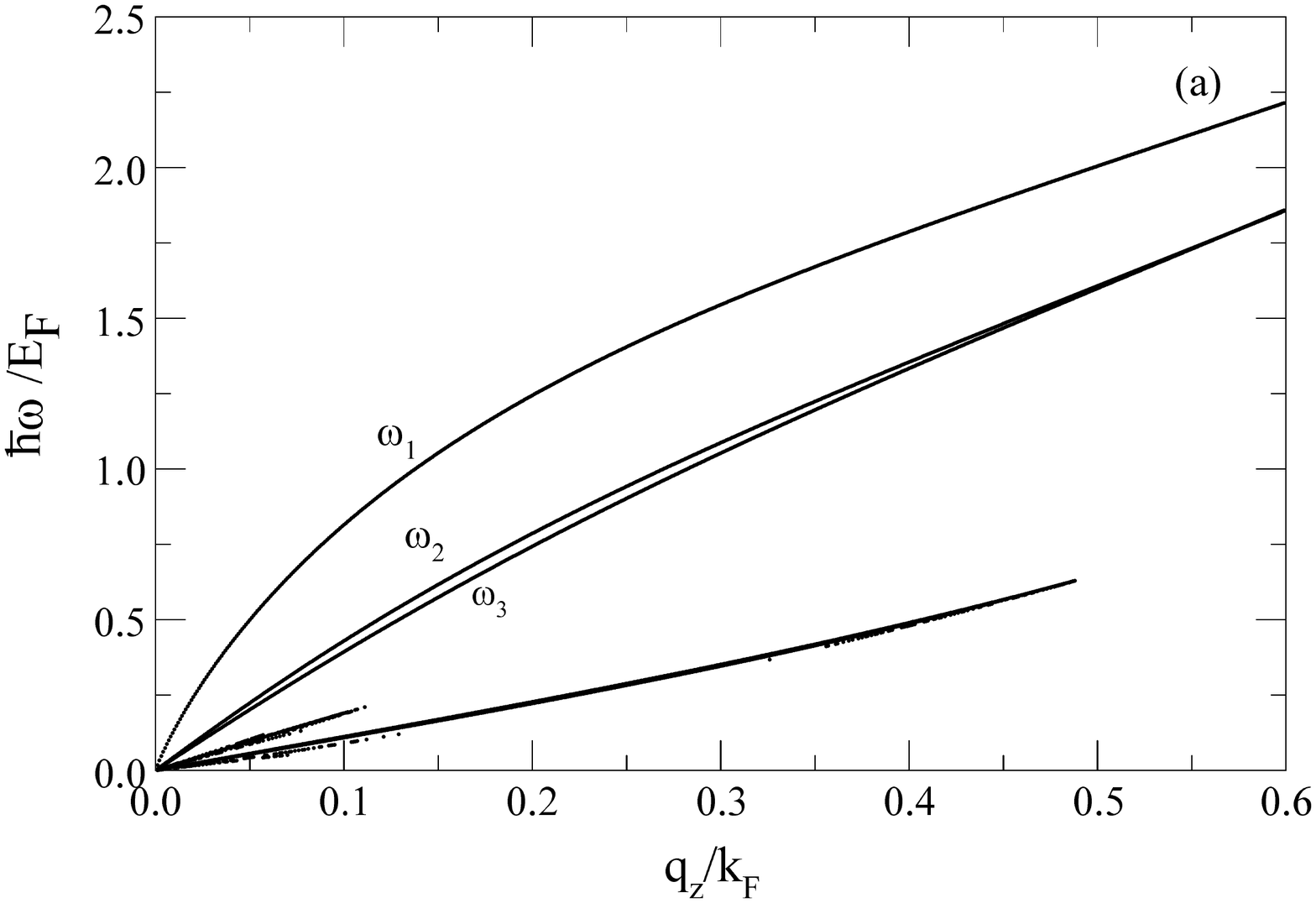}}}%
\qquad
\subfloat{%
\resizebox{4.5in}{3.5in}{\includegraphics{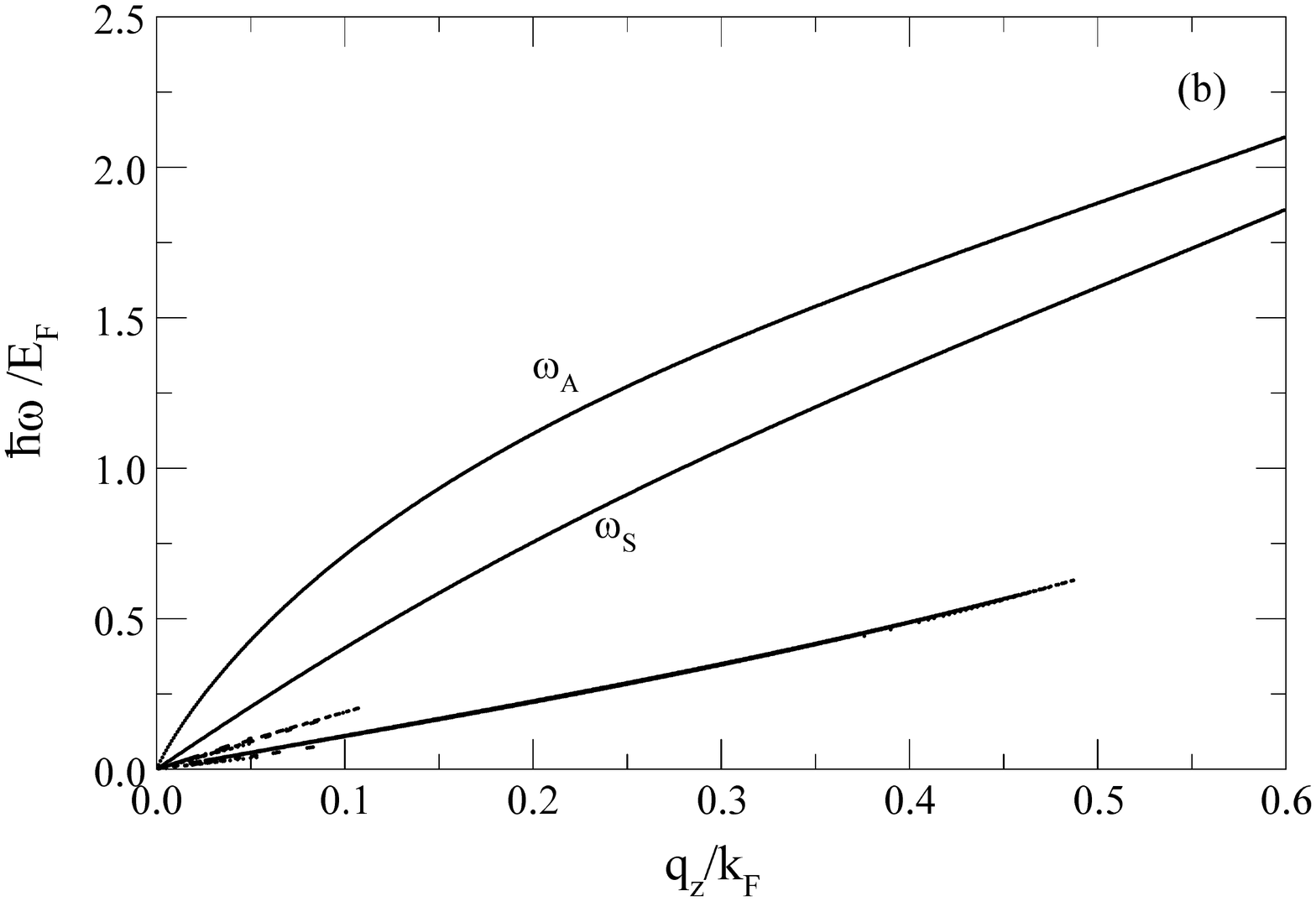}}}%
\caption{The intrasubband plasmon excitation energy, in units of the Fermi
energy $E_F$, as a function of $q_z/k_F$,   for the three singlewall nanotubes
described in Sec.~\ref{sec2}. The radius of each tubule was chosen as $11$
{\AA} and the centers of the tubules are at $a_x=25$ {\AA} and $a_y=25$ {\AA}
in (a). In (b), there are only two nanotubes with one fixed at the origin while
the position of the second nanotube on the $x$-axis is  at $a_x=25$ {\AA}.  All
other parameters are the
 same as in Fig.~\ref{fig2}. }
\label{fig7}
\end{figure}

\begin{figure}[p]
\centering
\subfloat{%
\resizebox{4.5in}{3.5in}{\includegraphics{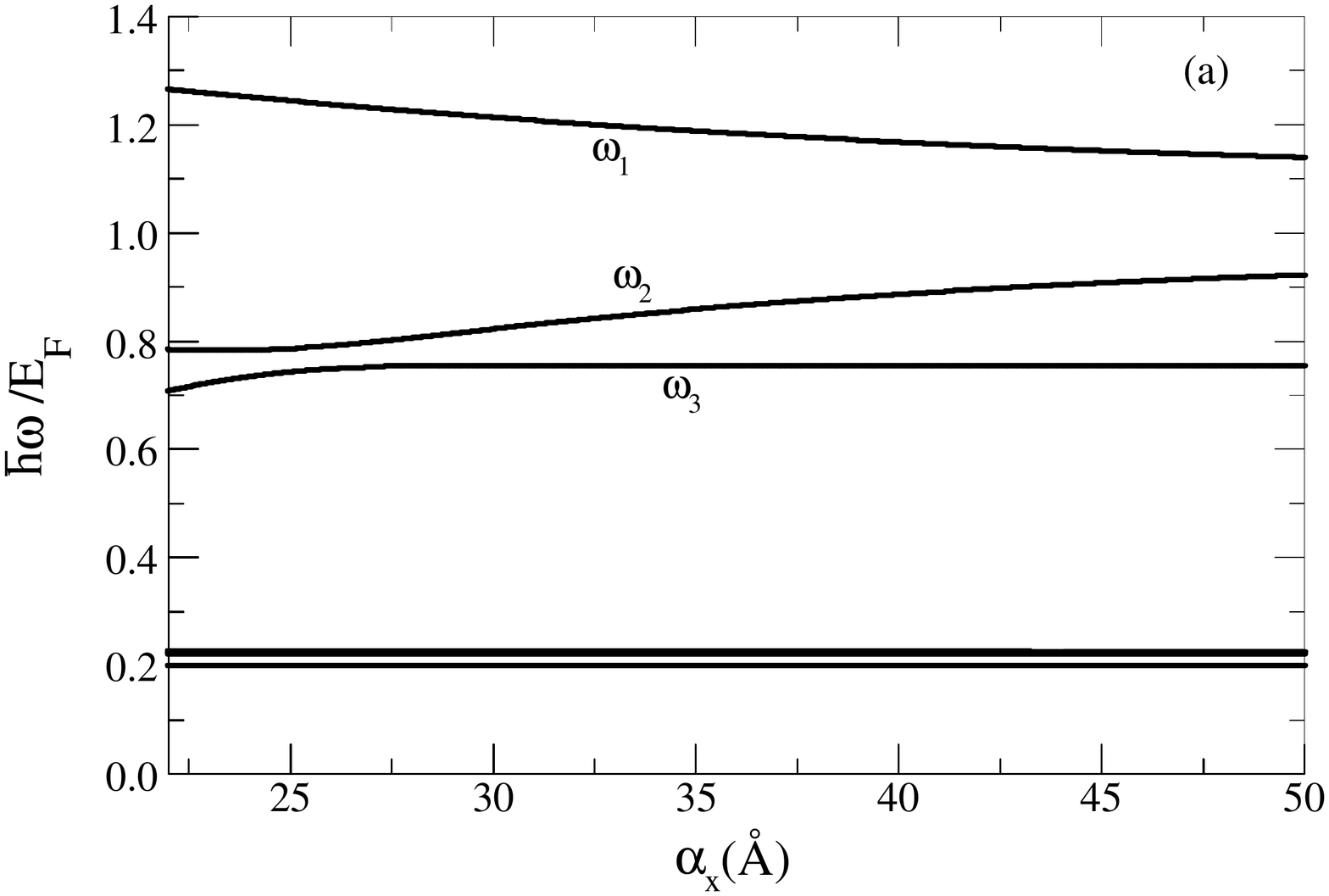}}}%
\qquad
\subfloat{%
\resizebox{4.5in}{3.5in}{\includegraphics{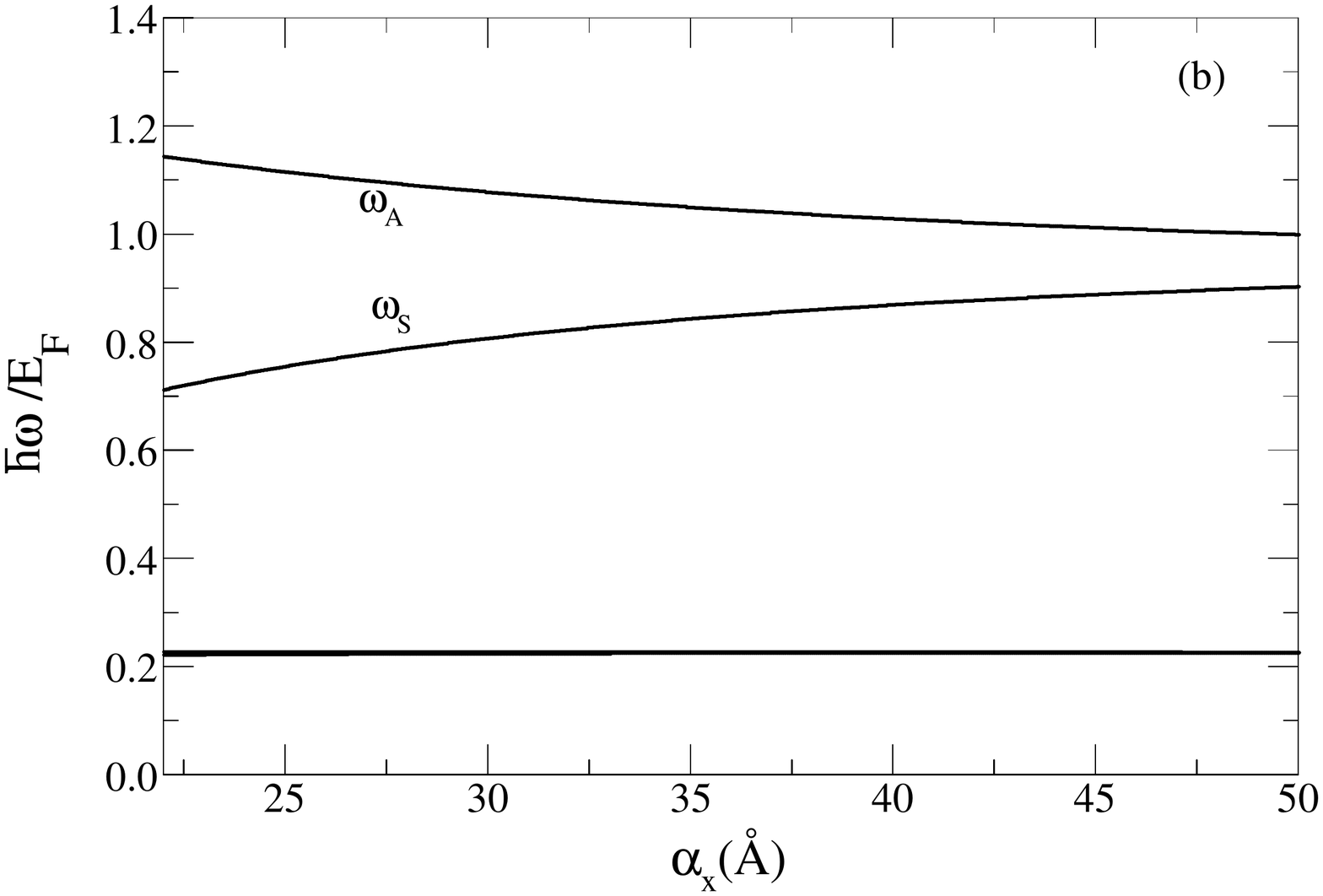}}}%
\caption{The intrasubband plasmon excitation energy, in units of the Fermi
energy $E_F$, as a function of $q_z/k_F$, for the three singlewall nanotubes
described in Sec.\ \ref{sec2}. The radius of each tubule was chosen as $11$ {
\AA} and the separation between the center of the tubules on the $y$-axis is
$a_y=25$ {\AA} in (a) while the  the coordinate of the third tubule on the
$x$-axis is varied. In (b), there are two nanotubes with one fixed at the
origin while the center of the second one on the $x$-axis is varied.  All other
parameters are the same as in Fig.~\ref{fig2}. }
\label{fig8}
\end{figure}

Figure \ref{fig7}(a)  shows the dispersion relation for intrasubband plasmon
excitations obtained by solving for the zeros of the determinant matrix in Eq.º
(\ref{5.1.211}). The plasmon energy is plotted as a function of the wave vector
$q_z$, for fixed $a_x=a_y=25.0$\ \AA. The radius of each of the three cylinders
is $11$\ \AA.  Each of the three cylinders has three plasmon branches when it
is not interacting with either one of the two other cylinders.    The Coulomb
interaction between electrons on different tubules splits the degenerate
plasmon modes for each tubule. The energy gap between split modes is very much
determined by the distance between the tubules. When $a_x$ or $a_y$ increases,
the separation between the split modes decreases, reducing to zero as the
limits $a_x,º a_y\to\infty$. Our numerical analysis shows that the amount by
which the branches is split depends on their frequency.  To understand the
results in Fig.\ \ref{fig7}(a)  better, we plot the dispersion relation for a
pair of singlewall nanotubes in Fig.\ \ref{fig7}(b)  with $a_x=25$\ \AA \ and all
other parameters  are the same as in Fig.\  \ref{fig7}(a). The two highest modes in
Fig.\  \ref{fig7}(b) are  symmetric ($\omega_S$) and antisymmetric ($\omega_A$)
modes. When the third nanotube is introduced in Fig.\  \ref{fig7}(a),  a third
high-frequency mode appears ($\omega_1,\omega_2$ and $\omega_3$) .   Since the
nanotube with its center located at $x=a_x$ is not separated by the same
distance from the naotube with its center at the origin and the nanotube with
its center at $y=a_y$, the plasmon modes $\omega_1,\omega_2$ and $\omega_3$ are
not equally spaced.  There is a total of nine plasmon modes in Fig.\
\ref{fig7}(a)  and six plasmon modes in Fig.\ \ref{fig7}(b), but not all of
them could be seen in these plots.  In Figs.\ \ref{fig8}(a)  and \ref{fig8}(b),
we fixed the value of the longitudinal wave vector at $q_z=0.2\ k_F$  and
plotted the intrasubband plasmon excitation energies as functions of $a_x$.
Figure \ref{fig8}(a)  was obtained for three nanotubes with $a_y=25$\ \AA \
whereas, Fig.\  \ref{fig8}(b)  was calculated for two nanotubes on the $x$-axis
with one of them fixed at the origin. The radius of each of the nanotubes was
chosen as $11$\ \AA\  and all other parameters are the same as in Figs.\
\ref{fig7}(a)  and  \ref{fig7}(b).    The highest plasmon modes are degenerate
in the limit $a_x\to \infty$ in Fig.\ \ref{fig8}(b)  for the pair of nanotubes.
The $\omega_1$  and $\omega_2$ modes become degeneate as $a_x\to \infty$. The
$\omega_3$ mode splits off from these two modes when the $a_x$ coordinate is
large due to the finite Coulomb interaction between the nanotubes on the
$y$-axis.  As a matter of fact, the $\omega_3$ branch is flat for large $a_x$
since the Coulomb interaction between nanotubes is not affected in this limit.
The low-frequency branches  shown in Figs.\  \ref{fig8}(a)  and \ref{fig8}(b)
clearly have a weak depence on the variation of the Coulomb interaction with
$a_x$.  When $a_x\approx 25$\ \AA, the $\omega_2$ and $\omega_3$ modes
anti-cross,  indicating that  beyond this distance the Coulomb interaction is
strongest between the two nanotubes on the $y$-axis.  In general, as shown in
Eq.º (\ref{mode-coupling}),  the modes with $m=0$ and $m\neq 0$ are coupled to
each  other.  To illustrate the way in which the Coulomb interaction modifies the plasmon modes for the three
 tubules, we only present results for $m=0$.   In a strict sense,
 the plasmon excitations cannot be categorized as intrasubband and intersubband plasmons
 as for a single tubule.

\section{Summary}
\label{sec5}

    In concluding this paper, we would like to point out that we used a simple electron gas model to obtain
the low-frequency plasmon excitation spectrum of a 2D periodic array  of parallel nanotubes and a bundle of three
nanotubes. The nanotubes are assumed doped   \cite{revi9}  in which the
charge carriers  are introduced onto the graphene tubules by means of intercalation,
which can be done in carbon fibers or $\rm C_{60}$  \cite{revi10}.
 Each nanotube consists of coaxial cylindrical tubules. The starting point of these calculations
was to model the electronic bandstructure of the tubules by a quasi-free electron gas confined to the surface
of an infinitely long cylinder of finite  radius. There was no tunneling of electrons between tubules.
 The random-phase approximation was employed to calculate the plasmon dispersion equation.
Plasmon excitation energies were obtained numerically for a single-wall nanotube array as a function of the
wave vector $q_z$ along the nanotube axes, with transverse wave vector $q_x$.  The Coulomb interaction between
nanotubes acts to split the single-nanotube plasmon modes.  We also obtained the periodic dependence of the
plasmon energies on $q_x$ reflecting the translational symmetry of the lattice. Our calculations should serve
as a framework for more elaborate computations using the tight-binding method for the
 bandstructure for the higher energy plasmon excitations on nanotubes. In our formalism, we showed that we
cannot be labeled by the  intrasubband or intersubband angular momentum
quantum numbers due to the Coulomb interaction between charge carriers on different tubules.

\acknowledgments

This work was supported by contract  FA 9453-07-C-0207
of AFRL.


\newpage

\end{document}